\documentclass[superscriptaddress,secnumarabic,amssymb,amsmath,nobibnotes,aps,prd,nofootinbib,showkeys,preprint]{revtex4}

\usepackage{graphicx}
\usepackage{float}
\usepackage{bm}
\usepackage{amsmath}
\usepackage{amsfonts}
\usepackage{amssymb}
\usepackage{epstopdf}
\usepackage{natbib}%
\newcommand{\bee}{\begin{equation}}
\newcommand{\eee}{\end{equation}}
\newcommand{\eaa}{\end{eqnarray}}
\newcommand{\baa}{\begin{eqnarray}}

\usepackage{color}

\begin{document}
\title{From modified Tsallis–R\'enyi entropy to a MOND-like force law, Bekenstein bound, and Landauer principle for black holes}


\author{Everton M. C. Abreu}
\email{ evertonabreu@ufrrj.br}
\affiliation{Physics Department, Universidade Federal Rural do Rio de Janeiro, RJ, Brazil}
\affiliation{Applied Physics Graduation Program, Physics Institute, Universidade Federal do Rio de Janeiro, RJ, Brazil}

\author{Jorge Ananias Neto}
\email{jorge.ananias@ufjf.br}
\affiliation{Physics Department, Universidade Federal de Juiz de Fora, Juiz de Fora, MG, Brazil}

\begin{abstract}
We examine black hole thermodynamics within the framework of modified R\'enyi entropy and explore its implications in Modified Newtonian 
Dynamics (MOND), an extension of Newton’s second law proposed to explain galaxy rotation curves without invoking dark matter. We conjecture 
that Tsallis entropy provides an exact description of  Bekenstein–Hawking entropy, from which the modified Rényi entropy is derived. Using 
this formulation, we show that a MOND-like force law emerges naturally from entropic considerations.
We also analyze the Bekenstein bound conjecture, which imposes an upper limit on the entropy of confined quantum systems, and verify its 
validity under the R\'enyi-modified framework for typical values of the deformation parameter $\lambda$. Furthermore, by invoking the 
Landauer principle, we obtain an expression for the mass loss due to black hole evaporation.
These results suggest that modified R\'enyi statistics, originating from Tsallis entropy, provides a coherent and promising approach 
to gravitational dynamics and information-theoretic aspects of black hole physics.
\end{abstract}


\maketitle
\section{Introduction}

The discovery that black holes (BHs) emit thermal radiation \cite{swh} was unexpected for much of the scientific community despite 
earlier indications of a fundamental connection between BH physics and thermodynamics. Bekenstein \cite{jdb} was among the first to 
identify thermodynamic properties in BHs, noting that their surface area behaves analogously to entropy.
According to Hawking’s area theorem \cite{swh}, the area $A$ of a BH never decreases in classical processes, mirroring the behavior of 
entropy in thermodynamic systems. This observation led to the recognition that the parallels between BH physics and thermodynamics are 
not merely coincidental but instead reveal a fundamental relationship between the two fields. Like conventional thermodynamic systems, 
BHs naturally evolve toward equilibrium, reaching a stable state after a characteristic relaxation timescale. Recent 
studies \cite{barrow, aab, aa1, aa2} have examined how a fractal structure in the BH horizon geometry affects its properties, further 
expanding the understanding of black hole thermodynamics.

The behavior of galaxy rotation curves is an important observational feature that has motivated the development of two competing models in the literature. 
The first model explains the observed discrepancies by postulating the existence of dark matter, whereas the second one suggests modifications to Newton’s 
fundamental laws of dynamics.
In this paper, we focus on the latter approach, commonly known as Modified Newtonian Dynamics (MOND) \cite{mm1,mm2,mm3}. However, MOND, 
as a fundamental theory, has known limitations that warrant consideration. Recent studies have questioned its presumed universality at 
galactic scales, suggesting that the constant \(a_0\) (see Eq. (\ref{mond}) below) may not be strictly universal, given the absence of a 
well-defined fundamental acceleration scale in galaxies \cite{Rodrigues:2018duc}. Furthermore, relativistic extensions of MOND, such as 
the TeVeS model, encounter significant difficulties in reproducing the cosmic microwave background (CMB) spectrum with 
precision \cite{Xu:2014doa}.

A key implication of the Landauer principle \cite{landauer} is that erasing a single bit of information in a computational system 
incurs a fundamental energy cost, resulting in heat dissipation. This energy cost is proportional to the system temperature and 
is fundamental to the thermodynamics of information processing. The principle stresses the physical limits of computation, 
reinforcing the fundamental link between information theory and thermodynamics.

Tsallis entropy belongs to a class of non-Gaussian entropies which have received significant attention in BH thermodynamics, 
as evidenced by the extensive research cited in references \cite{ci,tc,bc,mzlgsj}. Notably, the authors of references \cite{ci,bc} have 
introduced a novel variant of 
R\'enyi entropy for BH horizons. It is achieved by treating BH entropy as a nonextensive Tsallis entropy and employing a logarithmic formula. 
This approach yields a dual Tsallis entropy whose nonextensive effects enable the stabilization of BHs.

It is important to clarify that the Tsallis entropy considered in this work differs from the Tsallis--Cirto formulation \cite{tc}, which is commonly 
used in black hole thermodynamics. Here, we adopt the original Tsallis entropy, defined in terms of microstate probabilities. In contrast, the 
Tsallis--Cirto approach introduces a direct nonextensive modification of the Bekenstein--Hawking entropy by postulating a power law relation with the 
horizon area, written as \( S_q \propto A^q \), where \( q \) is the nonextensivity parameter. This difference has direct implications for black hole 
thermodynamics.

This paper explores the application of Tsallis’ modified entropy, which, as we will see, coincides with the modified R\'enyi entropy, 
in three distinct contexts.
First, we developed a modified gravitational force derived from the concept of entropic force, where holographic surface entropy 
plays an important role. We showed that the resulting effective gravitational force shares key similarities with the MOND framework.
Second, we demonstrated that Tsallis' modified entropy satisfies the Bekenstein bound conjecture \cite{beke}, a fundamental principle 
establishing an upper limit on entropy.
Finally, in the context of information theory, particularly Landauer’s principle, we calculated the mass loss of a black hole resulting 
from the erasure of a single bit of information. It is important to note here that a study of the modified Kaniadakis entropy has already 
been conducted \cite{mke}, from which a MOND-like theory was also derived, and the Bekenstein bound conjecture is satisfied within this framework.

We will show that the results obtained in this work using the modified R\'enyi entropy exhibit significantly different features compared to those
obtained with the modified Kaniadakis entropy \cite{mke}. 
This alone provides sufficient justification for a dedicated study on the application of the modified R\'enyi entropy. As we shall see in the 
section on Tsallis statistics, the modified R\'enyi entropy is closely related to the Tsallis entropy.

Before discussing Tsallis statistics and its connection to black hole thermodynamics through the modified Rényi entropy, we 
briefly note that replacing the Bekenstein--Hawking entropy \( S_{\text{BH}} \) with generalized formulations remains a subject of debate. 
Refs.~\cite{nof,gs} point to potential inconsistencies with thermodynamic laws and temperature definitions, and Ref.~\cite{pc} 
raises concerns about the use of Tsallis statistics in gravitational systems. Here, we adopt a framework where the modified Rényi 
entropy is viewed as a deformation of \( S_{\text{BH}} \) that preserves essential thermodynamic properties, ensuring stability, 
respecting the Bekenstein bound, and incorporating the Landauer principle into black hole evaporation. The significance of 
generalized entropy frameworks in gravitational physics remains an open and important question.

Based on these discussions, the structure of the paper is as follows: Section 2 reviews Tsallis thermostatistics. Section 3 
describes the main elements of the MOND mimicking framework. 
Section 4 analyzes the Bekenstein bound conjecture, and Section 5 discusses the Landauer principle. 
The conclusions, perspectives, and final comments are written in Section 6.

\section{Tsallis statistics}

Tsallis statistics \cite{t1,t2,t3} offer a non-extensive generalization for the standard Boltzmann-Gibbs (BG) 
statistics. Tsallis thermostatistics introduces a parameter $q$ modifying the usual entropy definition. More precisely,  
the Tsallis entropy is defined as
\begin{eqnarray}
S_q = k_B \, \frac{1 - \sum_{i=1}^{i=W} p_i^q}{q - 1} \,,
\end{eqnarray}
where $k_B$  is the Boltzmann constant, $p_i$ are the probabilities associated with each microstate, $q$ is the non-extensive parameter 
and $W$ is the total number of microstates.
This equation recovers the Boltzmann-Gibbs (BG) entropy when $q$ approaches 0. Remarkably, Tsallis-entropy exhibits the main 
properties of an entropy except for additivity. It is actually found that Tsallis entropy satisfies a  pseudo-additivity property.
This approach has been successfully used in many different physical systems. For instance, we can mention the Levy-type anomalous 
diffusion \cite{t4}, 
turbulence in a pure-electron plasma \cite{t5}, gravitational systems \cite{t6,t7,t8,t9}  and astrophysics \cite {adagp,piz,jsa,tj}.
It is noteworthy to affirm that Tsallis thermostatistics 
formalism has the BG statistics as a particular case in the limit q →1 where the standard additivity of entropy can be recovered.

Within the microcanonical ensemble, where all states have equal probability, Tsallis entropy simplifies to
\begin{eqnarray}
\label{km}
S_q = k_B\,\frac{W^{1-q} - 1} {1-q} \,.
\end{eqnarray}
This relation reproduces the standard BG entropy formula $ S = k_B \,\ln W$ in the limit $q \rightarrow 1$.

Now we propose that the Tsallis entropy, Eq. (\ref{km}), can describe the BH entropy, $S_{BH}$, 
from the equality
\begin{eqnarray}
\label{kbh} 
  k_B\,\frac{W^{1-q} - 1} {1-q}   = S_{BH}\,.
\end{eqnarray}
Therefore, from Eq. (\ref{kbh}), we have
\begin{eqnarray}
\label{w}
W = \left( 1 +  \lambda \frac{S_{BH}}{k_B} \right)^\frac{1}{\lambda} \,,
\end{eqnarray}
where $\lambda \equiv 1-q$. Using Eq. (\ref{w}) with the BG entropy, we obtain
\begin{eqnarray}
\label{kmod}
S_R = \frac{k_B}{\lambda}  \, \ln \left( 1 +  \lambda \,  \frac{S_{BH}}{k_B} \right) \,.      
\end{eqnarray}
Thus, Eq. (\ref{kmod}) represents a deformed version of the Rényi entropy, a previously known model (see, for example, Ref. \cite{ci}), 
which we will 
refer to here as the {\it modified R\'enyi entropy}.
Note that, when we assume $\lambda = 0$ in Eq. (\ref{kmod}), $S_R$ becomes $S_{BH}$. 
We can express the entropy $S_{BH}$ in Eq. (\ref{kmod}) as
\begin{eqnarray}
\label{sbh}
S_{BH} = \frac{k_B c^3 A}{4 G \hbar}     \,,     
\end{eqnarray}
where $c$ is the speed of light, $G$ is the gravitational constant, $\hbar$ is the Planck constant and $A$ is the area of the event horizon.
Using the relations \( \frac{G\hbar}{c^3} = l_p^2 \) and \( A = 4\pi R^2 \), the Bekenstein--Hawking entropy, Eq.~(\ref{sbh}), can be written as a 
function of the Planck length $l_p$ and the horizon radius \( R \) as
\begin{eqnarray}
\label{sbhr}
S_{BH} =  \frac{k_B A}{4 l_p^2} = \frac{k_B \pi R^2}{l_p^2} \,.
\end{eqnarray}

By differentiating the modified Rényi entropy, Eq.~(\ref{kmod}), with respect to the black hole mass, and assuming that the Bekenstein--Hawking 
entropy can be expressed as a function of the mass, given by
\begin{eqnarray} 
S_{\text{BH}} = \frac{4 \pi k_B G M^2}{\hbar c} \,,
\end{eqnarray} 
we obtain the corresponding Hawking temperature
\begin{eqnarray}
\label{htr} 
T = \left( \frac{dS_R}{dM} \right)^{-1} = \frac{\hbar c}{8 \pi k_B G M} + \frac{\lambda M}{2 k_B} \,.
\end{eqnarray}
In the limit \( \lambda \to 0 \), this expression reduces to the standard Hawking temperature, \( T = \hbar c / (8\pi k_B G M) \). 
Compared to the Bekenstein--Hawking result, the modified expression exhibits a minimum at a finite mass. As the mass decreases further, 
the temperature increases and diverges as \( M \to 0 \), suggesting that the black hole evaporates completely. Despite the deformation 
introduced by the parameter \( \lambda \), the qualitative features of the evaporation process remain essentially the same as in the 
standard case. 

It is worth noting that the Hawking temperature given in Eq.~(\ref{htr}) leads to thermodynamic stability in the sense that, beyond the critical 
mass of the black hole, a phase transition occurs at which the heat capacity becomes positive. This implies that the Schwarzschild black hole, when 
described using a deformable entropy \( S_{BH} \), specifically the modified Rényi entropy introduced in Eq.~(\ref{kmod}), becomes stable under thermal 
fluctuations. In contrast, the standard formulation consistently yields a negative heat capacity and hence instability. This distinction suggests that 
the modified Rényi entropy may offer a more consistent thermodynamic description in this context. Within this framework, the laws of black hole thermodynamics 
can be reformulated to incorporate the generalized entropy. For further details, see Refs.~\cite{ci,awn}.

\section{MOND mimicking theory}

The success of MOND theory stems from its effectiveness as a phenomenological model in describing the rotation curves of most galaxies. 
It successfully reproduces the well-known Tully-Fisher relation \cite{tf} and serves as a viable alternative to the dark matter paradigm. 
However, MOND theory has notable limitations, particularly in explaining the temperature distribution in galaxy clusters and 
addressing certain cosmological phenomena (see references \cite{anep, cs, eumond} for a detailed discussion). At its core, this theory 
modifies Newton’s second law, redefining the force in a way that can be expressed as
\begin{eqnarray}
\label{mond}
F = m \mu\left(\frac{a}{a_0}\right) \, a \,,
\end{eqnarray}
with $a$ denoting the usual acceleration, $a_0$ a suitable constant and $\mu(x)$ a real function possessing certain characteristics
\begin{eqnarray}
\label{mu1}
\mu(x)\approx  1 \;\; \mbox{for} \;\;  x >> 1 \,,
\end{eqnarray}
and
\begin{eqnarray}
\label{mu2}
 \mu(x) \approx x \;\; \mbox{for} \;\; x << 1\,.
\end{eqnarray}
Several functional forms of $\mu(x)$ have been proposed in the literature \cite{fgb,zf}. However, it is widely recognized that the core predictions 
of MOND theory remain largely independent of their exact mathematical formulation.

To derive a specific form of $\mu(x)$ from entropic principles, we begin with a general expression for the force, which is governed 
by the thermodynamic equation, as given in \cite{mondnos,modran}
\begin{eqnarray}
\label{nl}
 F =  \frac{G M m}{R^2} \, \frac{4l_p^2}{k_B} \,\frac{dS}{dA} \,,
\end{eqnarray}
where $A$ is the area of the holographic screen and $S$ is the entropy describing it.
Note that Eq. (\ref{nl}) is general enough to allow for the consideration of any type of entropy. For example,
if we consider the Bekenstein-Hawking entropy law, $ S_{BH} = k_B \,\frac{A}{4l_p^2}$,
as the entropy $S$ in Eq. (\ref{nl}), then                        
we can recover the usual Newton's universal law $ F = G M m/R^2$.
So, in order to derive a new effective gravitational force law in the context of R\'enyi modified entropy, initially we have

\begin{eqnarray}
\label{dsa}
\frac{dS_R}{d A} = \frac{  \frac{  dS_{BH}  }{   dA }   }{ 1 + \lambda \frac{S_{BH}}{k_B} } \,.
\end{eqnarray} 
Combining  Eqs. (\ref{sbhr}), (\ref{nl}) and (\ref{dsa}), we obtain a new effective gravitational force law as

\begin{eqnarray}
 \label{fmond}
 F_{effective} = \frac{ \frac{G M m}{R^2}  } {  1 +  \frac{\lambda \pi R^2}{ l_p^2 }  } \,.
\end{eqnarray}
We observe that, unlike the standard MOND theory, where the decay follows a $1/R$ behavior, Eq. (\ref{fmond}) exhibits a 
$1/R^4$ decay for large $R$. After some algebra, we obtain

\begin{eqnarray}
 \label{fmond2}
 F_{effective} = \frac{ m \, a_N}{  1 +  \frac{ a_0}{  a_N}  }  \,,
\end{eqnarray}
where
\begin{eqnarray}
 \label{ac}
 a_N \equiv \frac{G M}{R^2}  \,,
\end{eqnarray}
and

\begin{eqnarray}
 \label{a0}
a_0 \equiv  \frac{\lambda \pi G M}{l_p^2} \,.
\end{eqnarray}
Thus, from the effective gravitational force, Eq. (\ref{fmond2}), we can identify
\begin{eqnarray}
 \label{inter}
 \mu \left( \frac{    a_N     }{a_0} \right) = \frac{1}{1 + \frac{a_0}{a_N}} = \frac{    a_N  }{a_0} \left( 1+ \frac{   a_N }{a_0} \right)^{- 1}   \,.
\end{eqnarray}
The interpolating function in Eq. (\ref{inter}) clearly satisfies conditions (\ref{mu1}) and (\ref{mu2}).
Thus, we have shown that incorporating Rényi’s modified entropy, Eq. (\ref{kmod}), together with the force expression in Eq. (\ref{nl}), 
leads to a model analogous to MOND theory. Notably, the resulting interpolating function in Eq. (\ref{inter}) exactly matches 
the simple interpolating function used in MOND theory. 

Here, we would like to point out that, in the context of the modified Kaniadakis entropy \cite{mke}, 
the resulting interpolating function corresponds to the standard interpolating function, which provides a smoother 
transition between the Newtonian and deep MOND 
regimes when compared to the simple interpolating function derived from the modified R\'enyi entropy.

In addition, some approaches introduce quantum corrections to the Bekenstein--Hawking entropy, resulting in modifications 
of the gravitational force law. For instance, Ref.~\cite{bm} derives a modified MOND acceleration equation inspired by the generalized 
uncertainty principle (GUP), which includes higher order terms that decay faster than \( 1/R^2 \) and therefore do not reproduce 
the characteristic MOND behavior, where the acceleration scales as \( 1/R \) in the deep infrared regime. It is also worth noting that 
our model, based on the modified Rényi entropy, leads to a \( 1/R^4 \) decay at large distances, as previously discussed, and thus 
does not reproduce MOND either.

\section{Bekenstein bound conjecture}

The Bekenstein-Hawking entropy formula states that the entropy of a BH is directly proportional to the area of its event horizon. 
This principle plays a crucial 
role in BH thermodynamics, defining the connection between entropy and the surface area of the horizon. First proposed by Jacob 
Bekenstein and later refined by 
Stephen Hawking, it serves as a key framework for analyzing the thermodynamic properties of BHs and their implications to fundamental 
physics.

In addition, Bekenstein established a universal upper bound on the entropy of a confined quantum system \cite{beke}, expressed by
\bee
\label{bekenstein1}
S \leq \frac{2 \,\pi\, k_B\, R\, E}{\hbar \,c} \,,
\eee
where \( S \) denotes entropy, \( E \) represents the total energy, and \( R \) denotes the radius of a sphere enclosing the system. This inequality, known as the Bekenstein bound conjecture, suggests that as the Planck constant \( \hbar \) approaches to zero, entropy becomes unbounded, indicating that, in the classical 
limit, a localized system has no upper bound on entropy. Notably, the absence of Newton’s gravitational constant \( G \) 
in this formulation suggests that 
the Bekenstein bound conjecture might extend beyond gravitational interactions.

Although some counter-examples have been proposed, which may indicate potential limitations \cite{unruh,page,unruh2}, there is 
substantial evidence supporting the validity 
of the Bekenstein bound conjecture \cite{beke2,beke3,beke5,beke4}. This conjecture has been rigorously proven using standard quantum 
mechanics and quantum field theory in 
flat spacetime \cite{cas}. Additionally, the generalized uncertainty principle has played a role in deriving the Bekenstein bound \cite{bg}. 
The conjecture
has broad applications, particularly in cosmology and quantum field theory. For further insights, refer to \cite{bf,fs}.

Moving forward, we adopt the natural units system, where
\bee
\label{uni}
k_B = \hbar = c = G = 1  \,.
\eee
For these values, Eq. (\ref{bekenstein1}) simplifies to
\bee
\label{bekenstein2}
S \leq 2 \,\pi\, R\, E \,.
\eee
To obtain a form compatible with BH physics, we begin with the Schwarzschild metric, given by
\bee
\label{sch}
ds^2 = \left( 1 - \frac{2 M}{R} \right) dt^2 - \left( 1 - \frac{2 M}{R} \right)^{-1} dr^2 - r^2 d \Omega^2 \,.
\eee
The solution for BHs leads to the singularity
\bee
\label{bhs}
R = 2 M \,.
\eee 
Assuming the radius $R$ of the enclosing sphere corresponds to the Schwarzschild radius and considering
$E=M$, Eq. (\ref{bekenstein2}) can be rewritten as
\bee
\label{bekenstein3}
S \leq \pi R^2 \,.
\eee
We observe that Eq. (\ref{bekenstein3}) reaches its maximum when $ S = \pi R^2 $ for a Schwarzschild BH, with entropy given by $ S_{BH} = A_H/4 $ 
where $ A_H $ represents the horizon area, equal to $ 4 \pi R^2 $.

Hence, considering that the modified R\'enyi entropy governs BH thermodynamics, Eq. (\ref{kmod}) with $S_{BH} = \pi R^2$ allows us 
to recast the Bekenstein bound in
Eq. (\ref{bekenstein3}) as
\begin{eqnarray}
\label{kbb2}
S_R  \, \leq \,  \frac{e^{\lambda \,S_R} - 1}{\lambda } \,,  
\end{eqnarray}
To verify the validity of inequality (\ref{kbb2}), we have plotted in Figure 1 the ratio $R_R$,
\begin{eqnarray}
\label{kbb3}
R_R  \equiv  \frac{ e^{\lambda \,S_R} - 1 }{  \lambda S_R}   \,,
\end{eqnarray}
for typical values $\lambda \geq 0$ and $S_R = 2$.
\begin{figure}[H]
	\centering
   \includegraphics[height=7 cm,width=10 cm]{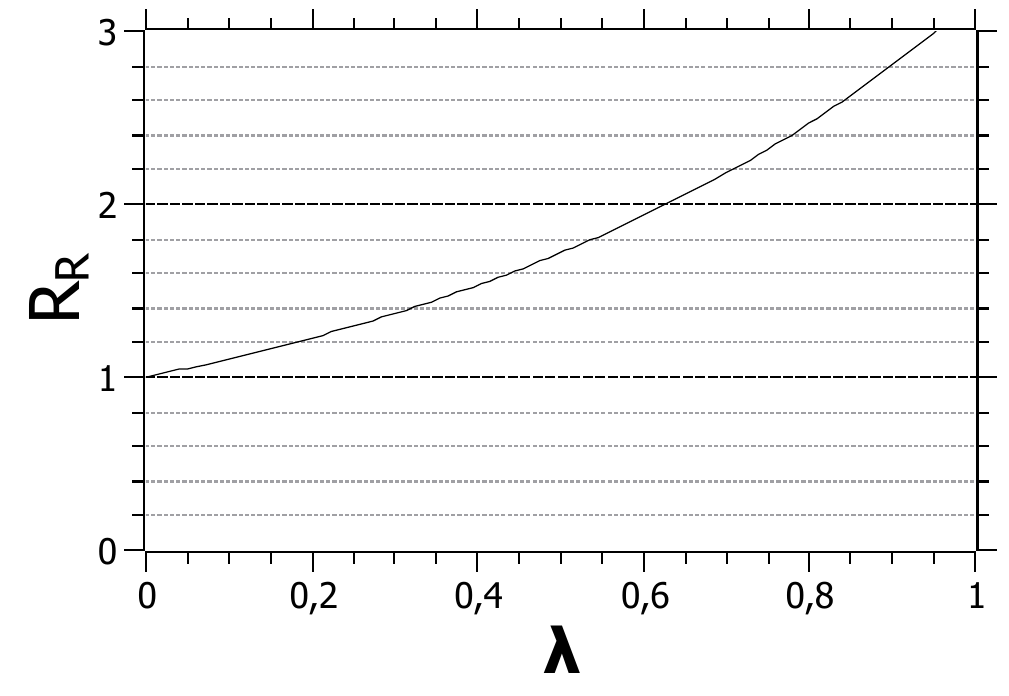}
	\caption{Values of $ R_R  \equiv  \frac{ e^{\lambda \,S_R} - 1 }{  \lambda S_R}   $ 
	as a function of $\lambda$ for $ S_R = 2$.}
	\label{kaniadakisb}
\end{figure}
In Figure 1, two key findings emerge. First, as $\lambda$ approaches 0, we observe that $R_R$ converges to 1. This result can 
be analytically verified by 
expanding the exponential function in a power series for small $\lambda$ values, yielding the approximation $\exp(\lambda S_R) \approx 1 + \lambda S_R$. 
Consequently, $R_R = 1$, confirming the equality stated in Eq. (\ref{kbb3}).
The second finding is that as the $\lambda$ parameter increases, $R_R$, which remains greater than one, also increases.
This result indicates that the Bekenstein bound conjecture holds when the modified Rényi entropy is applied to BH thermodynamics.

It is worth mentioning that both the Kaniadakis \cite{mke} and Rényi frameworks fully satisfy the Bekenstein bound. The difference lies in how 
strongly the bound is satisfied as the deformation parameters increase. For Kaniadakis entropy, the curve grows slowly and steadily, 
showing a mild and controlled deviation from the bound. In contrast, the R\'enyi entropy curve grows much faster, even for small parameter 
values, due to its exponential nature. Although the bound is still respected, R\'enyi entropy remains well below the theoretical limit, 
leaving a wider margin. This contrast illustrates how each model responds differently to the same fundamental constraint.

Here, it is essential to mention that we have shown that the usual Tsallis and Kaniadakis entropies and the Barrow entropy do not 
respect the Bekenstein bound conjecture. For further details, see reference \cite{euev}.

Before concluding this section, we note that our analysis of the Bekenstein bound relies on the standard identification \( E = M \),
valid for Schwarzschild black holes when natural units are adopted. Ref.~\cite{lgo} proposes a framework where the gravitational constant is modified 
and the energy appearing in the entropy bound is no longer identified with the black hole mass parameter. Although that work does not explicitly 
address the Bekenstein bound, this alternative formulation could, in principle, lead to different conclusions. In particular, generalized entropies 
such as those of Tsallis, Kaniadakis, and Barrow might be accommodated within this framework in a way that preserves the validity of the Bekenstein 
bound. In this paper, we adopt the conventional identification but recognize the potential relevance of this alternative perspective.

\section{Landauer principle}

The Landauer principle\footnote{Also known as the Brillouin principle \cite{bri}.}, proposed by Rolf Landauer in 1961 \cite{landauer}, 
is a fundamental concept linking information theory and thermodynamics. It asserts that the erasure of a single bit of information 
entails a minimum energy cost, given by the equation
\begin{eqnarray}
 \label{land}
 \Delta E \geq k_B T \ln 2 \,,
\end{eqnarray}
where $T$ signifies the absolute temperature. The $\ln 2$ term in Eq. (\ref{land}) appears 
because, in an information system, entropy, which measures the system level of disorder, is directly related to the number of 
possible microstates. This number is expressed as $\Omega = 2^N$, where $N$ denotes the number of bits.

The energy specified in Eq. (\ref{land}) is dissipated as heat into the surrounding environment. The Landauer principle stresses that 
information has a physical nature and it is fundamentally connected to thermodynamics. As a result, every operation that processes information 
is governed by thermodynamic principles.
Logical irreversibility, which refers to a computational process that can not be uniquely reversed, leads to physical irreversibility 
and contributes to an increase in entropy. Consequently, the erasure of information inevitably generates heat, in accordance with the second 
law of thermodynamics, which states that the entropy of 
an isolated system can not decrease. 

Here, it is important to emphasize that the original formulation of the Landauer principle does not explicitly incorporate 
gravitational effects. Nevertheless, our goal is to explore its applicability to gravitational systems, particularly in the context of black hole 
physics. Several studies have shown that the principle can be consistently extended to such regimes.

In a similar spirit, the Bekenstein bound conjecture, which we have employed in our analysis, does not involve the gravitational constant. 
Although originally motivated by black hole physics, it was formulated for general physical systems. These examples suggest that information 
theoretic principles may retain their validity even in the presence of gravity. In particular, the analysis presented in Ref.~\cite{lh} supports 
the view that the Landauer principle remains conceptually sound when applied to gravitational systems. For comprehensive reviews, 
see Refs.~\cite{plenio,frank}.

In the pioneering paper \cite{cl}, Landauer's principle was introduced into BH physics within the framework of Hawking evaporation. 
The authors of \cite{cl} demonstrated that Hawking evaporation fully complies with Landauer's principle, meaning that the information loss 
during the BH evaporation process occurs with maximum efficiency.
Subsequently, two recent papers \cite{otr, bgs} have extended the application of Landauer’s principle to new contexts: one addressing 
its implications for the apparent cosmological horizon \cite{otr} and the other exploring its role in the quantization of BH 
area \cite{bgs}. It is also worth noting that, prior to these developments, Abreu had already linked Hawking temperature to Landauer’s 
principle in \cite{abreu}.

In this approach, Landauer's principle can be derived directly from the Bekenstein-Hawking entropy law. Working in natural units 
where $\hbar = c = G = 1$, we recall that the entropy of a BH as a function of its mass is expressed as

\begin{eqnarray} 
\label{bhe} 
S = 4 \pi k_B  M^2 \,.
\end{eqnarray} 
Differentiating Eq.~(\ref{bhe}), the corresponding temperature of the black hole is given by

\begin{eqnarray} 
\label{bht} 
\frac{dS}{dM} = \frac{1}{T} = 8 \pi k_B  M \,. 
\end{eqnarray} 
Now, consider the scenario where the black hole loses enough mass to decrease its information content by one bit. 
Using Eq.~(\ref{bhe}), the associated entropy variation is

\begin{eqnarray} 
\label{dm} 
\Delta S = 8 \pi k_B  M \Delta M \,. 
\end{eqnarray} 
By assuming that the entropy change is given by

\begin{eqnarray} 
\label{dsb} \Delta S = k_B \ln 2 \,, 
\end{eqnarray} 
we can rewrite Eq.~(\ref{dm}) as

\begin{eqnarray} 
\label{ldm} \Delta M = \frac{\ln 2}{8 \pi } \frac{1}{M}\,. 
\end{eqnarray} 
Finally, substituting Eq.~(\ref{bht}) into Eq.~(\ref{ldm}) results in

\begin{eqnarray} 
\label{lds} \Delta M = k_B T \ln 2 \,. 
\end{eqnarray} 
Eq. (\ref{lds}) represents the saturated form of Landauer's principle. This confirms that the process of information loss during 
BH evaporation operates with maximal efficiency.

Our objective now is to determine the mass loss associated with BH evaporation under the framework of R\'enyi modified entropy. 
To achieve this, we start by computing the entropy variation, $\Delta S$, using Eq.~(\ref{kmod}), where $S_{BH}$ is defined by 
Eq.~(\ref{bhe}), leading to
\begin{eqnarray} 
\label{dmk} 
\Delta S = \frac{8 \pi  k_B  M \Delta M} { 1 + 4 \pi \lambda M^2} \,. 
\end{eqnarray} 
Using Eqs. (\ref{dsb}) in (\ref{dmk}), we then obtain
\begin{eqnarray} 
\label{dmassk} 
\Delta M =  \frac{\ln 2}{8 \pi } \, \frac{ 1 + 4 \pi \lambda M^2}{M} \,. 
\end{eqnarray}
By setting $\lambda = 0$ into Eq.~(\ref{dmassk}), we observe that Eq.~(\ref{ldm}) is recovered. 
In Fig.~2, we present the plot of the normalized mass variation, as given by Eq.~(\ref{dmassk}), as a function of mass $M$, considering 
$\lambda = 0$ (blue dashed line) and $\lambda = 0.2$ (red solid line).
\begin{figure}[H]
	\centering
    \includegraphics[height= 7 cm,width= 10 cm]{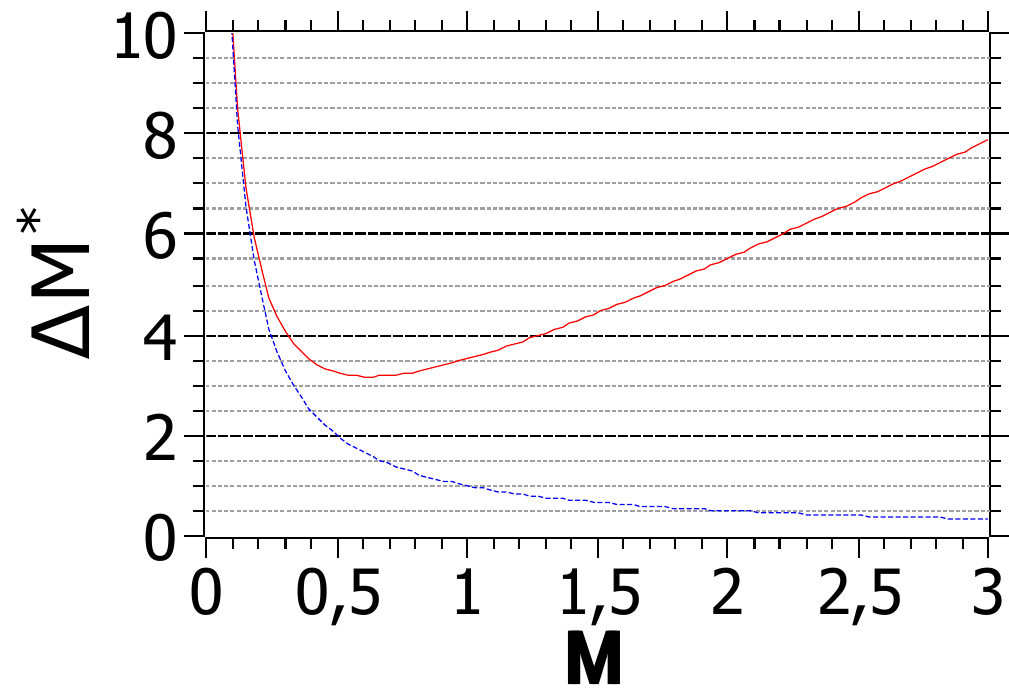}
	\caption{Values of the normalized black hole mass loss $ M^* \equiv 8 \pi {\Delta M}/{\ln 2} \, =\, 
          \frac{ 1 + 4 \pi \lambda M^2}{M} \,$ 
	as a function of $M$.}
	\label{landfig}
\end{figure}
From Fig.~2, we identify a minimum value for the BH mass loss, $\Delta M^*$, as described by Eq.~(\ref{dmassk}), 
for $\lambda = 0.2$ (red line). 
Beyond this minimum, it is evident that as $M$ increases, $\Delta M^*$ also grows, which contrasts with the behavior observed in Eq.~(32), 
where an increase in $M$ leads to a decrease in $\Delta M$. 
This result highlights a significant effect arising from the application of the modified R\'enyi entropy within Landauer's principle, 
particularly in the context of describing BH Hawking evaporation.

\section{Conclusions and perspectives}

In this paper, we have employed a modified R\'enyi entropy, initially proposed as a mechanism for ensuring the thermodynamic stability 
of BHs \cite{ci}. This approach is grounded in mathematical relationships linking different entropy formulations, such as 
Eq.~(\ref{kbh}), which expresses a connection between the number of microstates $W$ and the BH entropy $S_{BH}$. The modified 
R\'enyi entropy introduces a generalization in statistical mechanics through the 
inclusion of a parameter $\lambda$.
The MOND theory, which provides an alternative explanation for galaxy rotation curves in contrast 
to dark matter models, suggests a modification of Newton’s second law. It incorporates a function $\mu(x)$ with specific properties 
designed to alter the gravitational force in low-acceleration regimes. In our analysis, we derived an interpolation function given 
by Eq.~(\ref{inter}), indicating that the effective force described by Eq.~(\ref{fmond2}) exhibits characteristics resembling MOND 
theory under certain conditions.

The Bekenstein bound conjecture postulates a universal upper limit on the entropy of a confined quantum system. This conjecture has 
been validated across multiple contexts, including BH thermodynamics, where it establishes a constraint on the maximum entropy 
a BH system can attain. Our analysis shows that under the modified R\'enyi entropy framework, the Bekenstein bound conjecture is
satisfied.

In the context of Landauer’s principle, our findings indicate that, when described by the modified R\'enyi entropy, BH mass 
loss during Hawking radiation emission can become proportional to its mass beyond a certain threshold.
Additionally, it is worth emphasizing the significance of a key study \cite{bmenin} that aims to integrate the Bekenstein bound 
conjecture with Landauer’s principle, proposing a unified theoretical framework to understand the fundamental limits of entropy and 
energy within black hole systems.

As a proposal for future work, the use of the modified Rényi entropy could be explored in the context of BH 
area quantization. Since the area spacing is closely related to quantum gravity and the microscopic structure of entropy, 
examining how the modified Rényi entropy influences this quantization, particularly in connection with the Landauer principle, 
could provide further insight into the quantum aspects of BH thermodynamics.

In addition, Ref. \cite{57} the authors developed generalized formulations, a three and a four parameter equations, encompassing several 
famous kinds of entropy formulas. As a perspective we can apply these generalized formulation within Landauer's principle.   

As another perspective, in \cite{58} we have proposed to extend 
the Bekenstein bound conjecture to higher dimension. In this way we can ask how to extend the Landauer concepts to higher dimensions 
and coupled it to what 
we saw here.   It was suggested that the Barrow BH would be a supermassive BH dwelling in the center of a galaxy \cite{an}.   
It would be interesting to 
analyze statistically using R\'enyi statistics explored here this new formulation of supermassive BHs. 
These topics will be addressed in future investigations.

\section{Acknowledgments}
We are grateful to the anonymous Referee for the valuable suggestions.
Jorge Ananias Neto thanks CNPq (Conselho Nacional de Desenvolvimento Cient\'ifico e Tecnol\'ogico) for partial financial 
support, CNPq-PQ, Grant number 305984/2023-3.


\section{References}

 
\begin{thebibliography}{52}

\bibitem{swh} S. W. Hawking, Commun. Math. Phys. 43 (1975) 199.

\bibitem{jdb} J. D. Bekenstein, Phys. Rev. D 7 (1973) 2333.

\bibitem{barrow} J. D. Barrow, Phys. Lett. B 808 (2020) 135643.

\bibitem{aab} E. M. C. Abreu, J. Ananias Neto and E. M. Barboza, EPL 130 (2020) 40005.

\bibitem{aa1} E. M. C. Abreu and J. Ananias Neto, Phys. Lett. B 807 (2020) 135602. 

\bibitem{aa2} E. M. C. Abreu and J. Ananias Neto, Eur. Phys. J. C 80 (2020) 776.

\bibitem{mm1} M. Milgrom, Astrophys. J. 270 (1983) 365.

\bibitem{mm2} M. Milgrom, Astrophys. J. 270 (1983) 371.

\bibitem{mm3} M. Milgrom, Astrophys. J. 270 (1983) 384.

\bibitem{Rodrigues:2018duc}
D.~C.~Rodrigues, V.~Marra, A.~del Popolo and Z.~Davari,
Nature Astron. 2 no.8 (2018) 668. 
 
\bibitem{Xu:2014doa}
X.~d.~Xu, B.~Wang and P.~Zhang, Phys. Rev. D 92 (2015) 083505.

\bibitem{landauer} R. Landauer, IBM J. Res. Dev. 5 (1961) 183.

\bibitem{ci} V. G. Czinner and H. Iguchi, Phys. Lett. B 752 (2016) 306.

\bibitem{tc} C. Tsallis and  L. J. L. Cirto, Eur. Phys. J. C 73 (2013) 2487.

\bibitem{bc}  T. S. Bir\'o and V. G. Czinner,  Phys. Lett. B 726 (2013) 861.

\bibitem{mzlgsj} H. Moradpour, A. H. Ziaie, Iarley P. Lobo, J. P. Morais Gra\c ca, U. K. Sharma and A. Sayahian Jahromi,
Mod. Phys. Lett. A 37 12 (2022)  2250076.

\bibitem{beke} J. D. Bekenstein, Phys. Rev. D 23 (1981) 287.

\bibitem{mke}  G. V. Ambrósio, M. S. Andrade, P. R. F. Alves, C. N. Costa, 
J. Ananias Neto and R. Thibes, ``Exploring modified Kaniadakis entropy: MOND theory and the Bekenstein bound conjecture",
arXiv: 2405.14799v3 [gr-qc].

\bibitem{nof}  S. Nojiri,  S. D. Odintsov and  V. Faraoni,  Int. J. Geom. Meth. Mod. Phys. Vol. 19, No. 13 (2022) 2250210.

\bibitem{gs}  H. Gohar and V.  Salzano, ``On the foundations of entropic cosmologies: inconsistencies, possible solutions and
 dead end signs", arXiv: 2307.01768v3.

\bibitem{pc} Pedro Pessoa and Bruno Arderucio Costa,  Entropy, 22 (2020) 17.

\bibitem{t1} C. Tsallis, J. Stat. Phys. 52 (1988) 479.

\bibitem{t2} C. Tsallis,``Introduction to Nonextensive Statistical Mechanics: Approaching a Complex World", Springer, 2009.

\bibitem{t3} C. Tsallis, Braz. J. Phys. 29 (1999) 1.

\bibitem{t4} P.A. Alemany, D.H. Zanette, Phys. Rev. Lett. 75 (1995) 366.

\bibitem{t5} C. Anteneodo, C. Tsallis, J. Mol. Liq. 71 (1997) 255.

\bibitem{t6} J. Ananias Neto, Physica A 391 (2012) 4320;
E.M.C. Abreu, J. Ananias Neto, A.C.R. Mendes, Wilson Oliveira, Physica A 392 (2013) 5154;
Rafael C. Nunes, Edésio M. Barboza Jr., E. M.C. Abreu, J. Ananias Neto, J. Cosmol. Astropart. Phys. 1608 (2016) 08,051;
E.M.C. Abreu, J. Ananias Neto, A.C.R. Mendes, R. M. de Paula, Chaos Solitons and Fractals 118 (2019) 307.

\bibitem{t7} A. Majhi, Phys. Lett. B 775 (2017) 32.

\bibitem{t8} M. Tavayef, A. Sheykhi, Kazuharu Bamba, H. Moradpour, Phys. Lett. B 781 (2018) 195.

\bibitem{t9} C. Tsallis, Chaos Solitons and Fractals 13 (2002) 371.

\bibitem{piz} G. P. Pavlos, A. C. Iliopoulos, G. N. Zastenker, L. M. Zelenyi, L. P. Karakatsanis, M. O. Riazantseva, 
M. N. Xenakis, E. G. Pavlos, Physica A 422 (2015) 113.

\bibitem{adagp} M. Abrahão, W. G. Dantas, R. M. de Almeida, D. R. Gratieri and T. J. P. Penna, `` Fingerprint of Tsallis 
statistics in cosmic ray showers", arXiv: 1606.03923v1 [hep-ph].
 
\bibitem{jsa}  J. S. Almeida, Universe 2022, 1, 0.

\bibitem{tj}  C. Tsallis and H. J. Jensen, Phys. Lett. B 861 (2025) 139238.

\bibitem{awn} T. Anusonthi, P. Wongjun and R. Nakarachinda, ``Thermodynamic Stability of Schwarzschild-de Sitter Black holes 
with Rényi entropy", arXiv: 2501.04378v1  [gr-qc].
 
\bibitem{tf} R. B. Tully and J. R. Fisher,  Astron. Astrophys. 54 (1977) 661.

\bibitem{anep} A. Nusser, E. Pointecouteau, Mon. Not. R. Astron. Soc. 366 (2006) 969.

\bibitem{cs} C. Skordis, Class. Quantum Gravity 26, 143001 (2009).

\bibitem{eumond} J. Ananias Neto, Int. Jour. Theor. Phys. 50 (2011) 3552.

\bibitem{fgb}  B. Famaey, G. Gentile and J. P. Bruneton,  Phys. Rev. D 75 (2007) 063002.

\bibitem{zf} H. S. Zhao and B. Famaey, J. Astrophys. L9 (2006) 638.

\bibitem{mondnos} E. M. C. Abreu, Jorge Ananias Neto, Albert C. R. Mendes and Daniel O. Souza, EPL 120 (2018) 20003.

\bibitem{modran} L. Modesto and A. Randono, ``Entropic Corrections to Newton's Law'', arXiv:1003.1998 [hep-th].

\bibitem{bm} B. Majumder, Adv. High Energy Phys., vol. 2013, article ID 296836.

\bibitem{unruh} W. G. Unruh and R. M. Wald, Phys. Rev. D 25 (1982) 942.

\bibitem{page} D. N. Page, Phys. Rev. D 26 (1982) 947.

\bibitem{unruh2} W. G. Unruh and R. M. Wald, Phys. Rev. D 27 (1983) 2271.


\bibitem{beke2} J. D. Bekenstein, Phys. Rev. Lett. 46 (1981) 623.

\bibitem{beke3} J. D. Bekenstein, Gen. Relativ. Gravit. 14 (1982) 355.

\bibitem{beke5} M. Schiffer and J. D. Bekenstein, Phys. Rev. D 39 (1989) 1109.

\bibitem{beke4} J. D. Bekenstein, Phys. Lett. B 481 (2000) 339.

\bibitem{cas} H. Casini, Class. Quant. Grav. 25 (2008) 205021.

\bibitem{bg} L. Buoninfante, G. G. Luciano, L. Petruzziello and F. Scardigli, Phys. Lett. B 824 (2022) 136818.

\bibitem{bf} T. Banks and W. Fischler, ``Cosmological Implications of the Bekenstein Bound", in Jacob Bekenstein (2019) 121 World Scientific.

\bibitem{fs} F. Scardigli, Universe 8 (2022)  645.

\bibitem{euev} E. M. C. Abreu and J. Ananias Neto, Phys. Lett. B 835 (2022) 137565.

\bibitem{lgo} H. Lu, S. Di  Gennaro  and Y. C. Ong, Annals Phys. 474 (2025) 169914.

\bibitem{bri} L. Brillouin, J. Appl. Phys. 24 (1953) 1152.

\bibitem{lh} L. Herrera, Entropy, 22 (2020) 340.

\bibitem{plenio} M. B. Plenio and V. Vitelli, Contemporary Physics 42 (2001) 25.

\bibitem{frank} M. P. Frank, ``Physical Foundations of Landauer’s Principle". In: Kari, J., Ulidowski, I. (eds) Reversible Computation. RC 2018. 
Lecture Notes in Computer Science, vol 11106.

\bibitem{cl} M. Cort\^es and A. Liddle, ``Hawking evaporation and the Landauer principle", arXiv: 2407.08777v2 [gr-qc].

\bibitem{otr} O. Trivedi, ``Universality of Information Thermodynamics and the Efficiency of Information Erasure on the 
Cosmological Apparent Horizon",  arXiv: 2407.15231v2 [gr-qc].

\bibitem{bgs} B. Bagchi, A. Ghosh and S. Sen, Gen. Relativ. Gravit. 56 (2024) 108.

\bibitem{abreu} E. M. C. Abreu, ``Barrow black hole variable parameter model connected to information theory",  
arXiv: 2402.15922v1 [gr-qc].

\bibitem{bmenin} B. Menin, Journal of Applied Mathematics and Physics 11 (2023) 2185.

\bibitem{57}  S. Nojiri, S. D. Odintsov and T. Paul, Phys. Lett. B 835 (2022) 137553.

\bibitem{58}  E.M.C. Abreu and J. Ananias Neto, ``The Bekenstein bound of Schwarzschild black hole in higher dimensions", submitted for publication.

\bibitem{an}   E. M. C. Abreu and M. J. Neves, Phys. Lett. B 864 (2025) 139431.

\end{thebibliography}
\end{document}